\documentclass[aoas,nameyear,MSNbibl,seceqn,dvips]{arximspdf}
\usepackage{dcolumn}
\usepackage{graphicx}

\doi{10.1214/10-AOAS400}
\volume{5}
\issue{2A}
\pubyear{2011}
\firstpage{645}
\lastpage{668}

\makeatletter

\newcolumntype{d}[1]{D{.}{.}{#1}}

\newcommand{\by}{\mathbf{y}}
\newcommand{\Exp}{\mathrm{E}}
\newcommand{\Expec}{\mathrm{E}}
\newcommand{\Prob}{\mathrm{P}}
\newcommand{\Var}{\operatorname{var}}
\newcommand{\Cov}{\operatorname{cov}}
\newcommand{\tell}{\tilde{\ell}}
\newcommand{\argmax}{\operatorname{argmax}}
\newcommand{\median}{\operatorname{median}}

\newproclaim{remark}{Remark}
\newproclaim{remarks}{Remarks}

\makeatother

\begin{document}
\begin{frontmatter}

\title{Detecting simultaneous variant intervals in~aligned sequences}
\runtitle{Simultaneous variant interval detection}

\begin{aug}
\author[A]{\fnms{David} \snm{Siegmund}\thanksref{t2}\ead[label=e1]{dos@stat.stanford.edu}},
\author[B]{\fnms{Benjamin} \snm{Yakir}\thanksref{t2}\ead[label=e2]{msby@mscc.huji.ac.il}}
and
\author[A]{\fnms{Nancy R.} \snm{Zhang}\corref{}\thanksref{t3}\ead[label=e3]{nzhang@stanford.edu}}
\thankstext{t2}{Supported in part by the Israeli-American Bi-National Fund.}
\thankstext{t3}{Supported in part by NSF DMS Grant ID 0906394.}
\runauthor{D. Siegmund, B. Yakir and N. R. Zhang}
\affiliation{Stanford University, Hebrew University of Jerusalem
and Stanford University}
\address[A]{D. Siegmund\\
N. R. Zhang\\
Department of Statistics\\
Stanford University\\
Sequoia Hall\\
390 Serra Mall\\
Stanford, California 94305-4065\\
USA\\
\printead{e1}\\
\hphantom{E-mail:\ }\printead*{e3}} 
\address[B]{B. Yakir\\
Department of Statistics\\
Hebrew University of Jerusalem\\
Jerusalem 91905\\
Israel\\
\printead{e2}}
\end{aug}

\received{\smonth{2} \syear{2010}}
\revised{\smonth{6} \syear{2010}}

%
\begin{abstract}
Given a set of aligned sequences of independent noisy observations,
we are concerned
with detecting intervals where the mean values of the observations
change simultaneously in a subset of the sequences.
The intervals of changed means are typically short relative to the
length of
the sequences, the subset where the change occurs, the ``carriers,''
can be relatively small, and the
sizes of the changes can vary from one sequence to another.
This problem is motivated by the scientific
problem of detecting inherited copy number variants in
aligned DNA samples. We suggest a statistic based on the
assumption that for any given interval of changed means there is
a given fraction of samples that carry the change.
We derive an analytic approximation for the false positive error
probability of a scan,
which is shown by simulations to be reasonably accurate. We show that
the new method usually improves on methods that analyze a
single sample at a time
and on our earlier multi-sample method, which is most efficient when
the carriers form a large fraction of the set of sequences.
The proposed procedure is also shown to be robust with
respect to the assumed fraction of carriers of the changes.
\end{abstract}

%
\end{frontmatter}

\section{Introduction}

This paper is motivated by the problem of detecting inherited
DNA copy number variants (CNV). CNV are gains and losses of
segments of chromosomes, and comprise an important class of
genetic variation in human populations. Various laboratory
techniques have been developed to measure DNA copy number
[\citet{Pinkel1988}; \citet{Pollack}; \citet{Snijders2001};
\citet{Bignell2004}; \citet{Peiffer2006}].
These measurements are taken at a set of probes, each mapping
to a specific location in the genome. The data thus produced
are a set of linear sequences of measurement intensities, one
for each biological sample in the study. If a sample contains
a CNV at a particular genomic region,
then depending on whether the CNV is a gain or loss,
the intensities increase
or decrease relative to their average values in that region.

\begin{figure}

\includegraphics{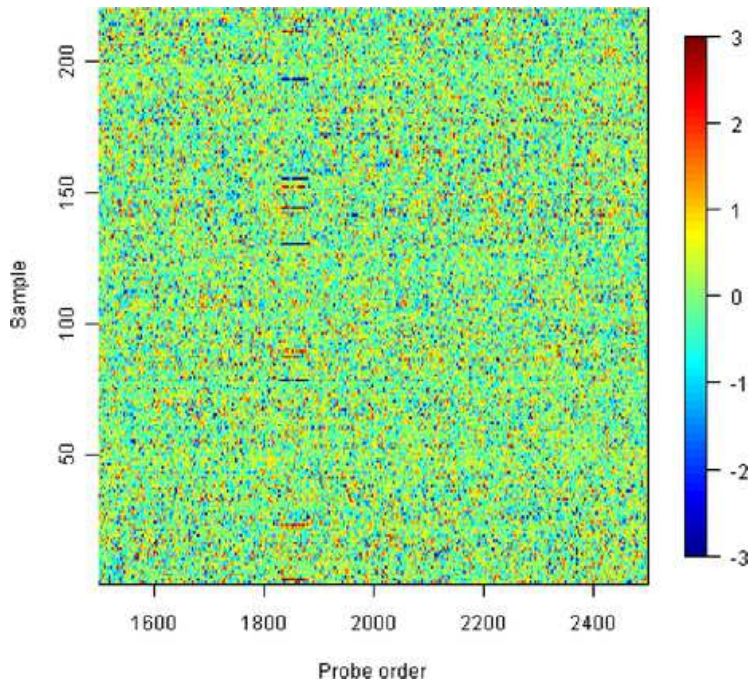}

\caption{An example segment of DNA copy number data. Each row is a
sample, and each column is a probe. Gains and losses in copy number
manifest as stretches of low or high intensities.}\label{fig:1}
\end{figure}

Studies of DNA copy number arise in two distinct contexts,
which yield data with different characteristics. One of these
is cancer genetics, where somatic changes in DNA copy number
occur in the genomes of tumor cells. [See
Pinkel and Albertson \citep{PinkelAlbertson} for a review.] These
changes can be
quite long, sometimes involving entire chromosomes or
chromosomal arms. The second context, which motivates the
problem formulation in this paper, involves inherited regions
of CNV. These are population polymorphisms. As such, they
hypothetically could be functional variants contributing to
phenotypic variability, and hence are
of interest in association studies. Alternatively, they can be neutral
markers for tracing
distant relationships in populations, which could be used in
population genetics. Since inherited regions
of CNV are typically quite short, often covering only one or a
few probes, they are more difficult to detect in individual
genomes than their tumor counterparts, which has led some
investigators to place a~minimal length of 2--10 probes on a
CNV [e.g., Redon et al. \citep{Redon}, McCarroll et al. \citep{McCarroll2008},
Walsh et al. \citep{walsh}] even though this restriction artificially
eliminates many
candidates from contention. An illustrative segment of CNV data
from a group of normal samples are shown in the form of a
heatmap in Figure \ref{fig:1}. Each row of the heatmap is a
sample, and each column is a probe. The probes map to ordered
locations along a chromosome. For illustration, the region
depicted in Figure \ref{fig:1} contains a CNV between probes
1800 and 1900 that is visibly apparent as stretches of high
(red) or low (blue) intensities in a few of the samples. Note
that the breakpoints are shared across samples, and that the
shift in mean may be positive for some individuals and negative
in others.

%

Most current procedures process the samples one at a time in
the detection of CNV. For recent reviews, see Lai et al. \citep
{LaiPark}, Willenbrock and Fridlyand
\citep{Willenbrock2005}, and Zhang \citep{Zhang2010a}.
Lai et al. and Willenbrock and Fridlyand compare many of the
existing methods on a common data set. In this paper we take
the view that since these CNV are population level
polymorphisms, there is the possibility to pool data across
individuals (samples) to boost the power of detection of
simultaneous changes occurring in a fraction of the sequences.
See Zhang et al. \citep{Zhang2009b} for more scientific background and
additional references.

Following Olshen et al. \citep{Olshen2004}, we formulate this problem
as one
of detecting intervals where the mean of a sequence of
independent random variables shows a change from its baseline, that is, overall
mean,
value. Zhang et al. \citep{Zhang2009b} extended the approach of Olshen
et al. to the case of multiple aligned
sequences and the
problem of detecting
intervals
of change that occur at identical locations in
some of the sequences. They proposed a sum of
chi-squares statistic, which is effectively the likelihood
ratio statistic assuming normal errors, and
showed that a simultaneous scan of all sequences for a shared signal
across multiple profiles can improve power compared to a method
that separately segments each individual sequence, especially
if a moderate to large fraction of the sequences
``carry'' the change. [The methods of Olshen et al. \citep{Olshen2004}
and Zhang et al. \citep{Zhang2009b} are reviewed in more detail in
Section \ref{sec2.2}.]

Since the sum of chi-squares statistic was designed for the
situation where a moderate to large fraction of the sequences
carry a change, it can have low power to detect the many CNV
that are \emph{rare variants}, where the fraction of carriers
is less than $ \sim$5\%. The accurate detection of rare
variants is becoming increasingly important, due to the recent
interest in association studies targeting rare variants [cf.
the review by McCarroll \citep{McCarroll2008c}]. Although
Zhang et al. \citep{Zhang2009b} also suggested a class of ``weighted''
statistics to detect rare variants, the method they used to
approximate $p$-values for the sum of chi-squares statistic
relies on the spherical symmetry of the standard multivariate
normal distribution, and does not adapt to the more general
scan statistics considered in this paper. Our main theoretical
result is a more general method to approximate the false
positive rate for a wide class of multi-sample scan statistics,
which includes the sum of chi-squares statistic as a special
case. We show by simulations that the approximations are quite
accurate. This allows us to assess the significance of
genome-wide studies, which often involve over a million probes
and thousands of samples. Simulations and other computer
intensive methods are very difficult to implement for scans of
such large data sets.

In Section \ref{sec:model} we formulate the basic model and
suggest a class of statistics based on the assumption of a
mixture of mean levels at each variant interval. Next we
generalize the method introduced by Siegmund, Yakir and Zhang
(\citeyear{SiegmundYakirZhang}) to provide analytic approximations to the false positive
rates of these statistics, and we use Monte Carlo experiments
to show that the approximations are very accurate. In Section
\ref{sec:power} we compare different statistics and illustrate
the benefits of pooling information across
samples, even in the case where the proportion
of carriers is very low. Section \ref{sec:data} contains a
test case involving actual CNV data. Section \ref{sec6}
contains a discussion, and in Appendix \ref{sec:proofs} we
sketch a proof of our false positive rate approximations.

The independence and normality assumptions made in this paper
also underlie most previous approaches to this problem. Raw
data from popular genotyping microarray platforms often deviate
from these assumptions, but most of this deviation can be
eliminated by appropriate normalization procedures. A description of
data preprocessing is given in Section
\ref{sec:data}.

We consider here the primary problem to be detection of the
intervals of change. In many cases, the carriers, that is, the
subset of samples where the changes have occurred, are
relatively obvious from inspection of the data after the intervals have
been reported. In other
cases, determining the carriers poses a difficult auxiliary
problem, because of the very large
dimension of the parameter space. Zhang et al. \citep{Zhang2009b}
suggested a
simple ad hoc
thresholding algorithm. We expect to
discuss in the future more systematic criteria that
involve modeling of probe-specific effects, clustering
across samples, and a generalization to multiple sequences
of the BIC method of Zhang and Siegmund \citep{Zhang2007}.

For data from some platforms (e.g., the SNP
genotyping arrays from Affymetrix and Illumina), other
information, such as A and B allele frequencies, is available to
improve the accuracy of CNV detection. Some methods
[\citet{PennCNV}; \citet{colella2007}] use a Hidden Markov
model to
detect CNV based on both the total intensity and the
allele specific data.
While
\citet{colella2007} mentioned that their hidden Markov
model can be extended to process multiple samples
simultaneously,
no convincing evidence was presented that the
allele specific analysis, when combined across samples, improves
accuracy. The reason, at least for the Affymetrix
platform, is that allele specific frequencies are also prone to
artifacts and
can be much noisier than total intensity data. While effective
measures for artifact removal for total intensity data have
been developed (see Section~\ref{sec:data}) and allow
successful cross-sample integration, appropriate measures appear to be
lacking for normalization of allele specific frequencies.
Although methods based on allele specific data undoubtedly have a
role to play in CNV detection,
in this paper we focus on the integration of total
intensity data across samples, which admits an appealingly
simple and general model that appears to be more generally useful.
%
%

\begin{remark*}
Although the formulation and
results in this paper have been motivated by problems
associated with detection of CNV, the multisample change-point model
that we
study may be useful in quite different contexts. One of
current interest is sequential detection of a change-point by
a distributed array of sensors [e.g., Tartakovsky and Polunchenko
\citep{TartakovskyPolunchenko}], where our $p$-value approximation
can be used as the starting point to develop an approximation to the average
run length when there is no change-point.
Another example is briefly described in the \hyperref[sec:proofs]{Appendix}.
\end{remark*}

\section{Change-point models and scan statistics}\label{sec:model}

\subsection{Problem formulation} The observed data is a two-dimensional array $\{y_{it}\dvtx   1\leq i \leq N,   1\leq t \leq
T\}$, where $y_{it}$ is the data point for the $i$th profile
at location $t$, $N$ is the total number of profiles, and $T$
is the total number of locations. In genome-wide profiling
studies, $N$ is usually in the tens to the thousands, and $T$ is
usually in the hundreds of thousands. We assume that for each
$i$, the random variables $\by_i=\{y_{it}\dvtx  1\leq t \leq T\}$
are mutually independent and Gaussian with mean values
$\mu_{it}$ and variances $\sigma_i^2$. Under the null
hypothesis, the means for each profile are identical across
locations. Under the alternative hypothesis of a single changed
interval, there exist values $1 \leq\tau_1 < \tau_2 \leq T$
and a set of profiles ${\mathcal{J}} \subset\{1,\ldots,N \}$, such
that for $i \in\mathcal{J}$,
%
\begin{equation} \label{model} \mu_{it} = \mu_{i} + \delta_i
I_{\{ \tau_1 < t \leq\tau_2\}},
\end{equation}
where the $\delta_i$ are nonzero constants and
$\mu_{i}$ is the baseline mean level for profile~$i$, which may
not necessarily be known in advance. Under the alternative
hypothesis we refer to $(\tau_1,\tau_2]$ as a variant interval
and $\mathcal{J}$ as the set of carriers, that is, the subset of samples
that have a changed mean in that interval. If the alternative
hypothesis is true, we are interested primarily in detecting
this situation and in estimating the endpoints of the variant
interval, and secondarily in determining the carriers.

In DNA copy number data, the magnitude of change in signal
intensity varies across samples
for any given CNV, even when the underlying change in copy number is
the same. This is due
to differences in sample handling, and motivates the
assignment of a new $\delta_i$ parameter to each carrier; see
Zhang et al. \citep{Zhang2009b} for examples.

In many applications, including CNV detection, there
are usually multiple variant
intervals defined by different $\tau_1$, $\tau_2$ and
$\mathcal{J}$. We describe the model and statistics assuming the
simple case where there is at most one variant interval.
If the number of intervals is small and the intervals are widely spaced,
a single
application will detect multiple intervals.
More generally, these
statistics can be combined with the recursive segmentation
algorithm in Zhang et al. \citep{Zhang2009b} to treat the case where there
are multiple variant intervals.

\subsection{Review of scan statistics}\label{sec2.2}

First we review the case of a single sequence of observations.
Initially we suppress the dependence of our notation on the
profile indicator $i$. For $\{y_1, \ldots, y_T\}$, let $S_t =
y_1 + \cdots+ y_t$, $\bar{y}_t = S_t/t$, and $\hat{\sigma}^2 =
T^{-1} \sum_1^T (y_t - \bar{y}_T)^2$ be the maximum likelihood
estimate of variance. Olshen et al. \citep{Olshen2004} used likelihood ratio
based statistics for analysis of DNA copy number data for a
single sequence. The statistic they suggested was
%
\begin{equation} \label{maxchisquare}
\max_{s,t} U^2(s,t),
\end{equation}
where
\begin{equation} \label{chgptstat} U(s,t)
=\hat{\sigma}^{-1} \{S_t - S_s - (t-s)\bar{y}_T\}/
[(t-s)\{1-(t-s)/T\}]^{1/2},
\end{equation}
and the max is taken over
$1 \leq s < t \leq T,    t-s \leq T_1$. Here $T_1 < T$ is an
assumed upper bound on the length of the variant interval, which
for some applications may be much smaller than $T$.

If the error standard deviation $\sigma$ were known and used in
place of $\hat{\sigma}$ in (\ref{chgptstat}),
 (\ref{maxchisquare}) would be the likelihood ratio statistic.
The denominator in (\ref{maxchisquare}) standardizes the
variance of the numerator, and under the null hypothesis of no
change, $U^2(s,t)$ is asymptotically distributed as $\chi^2_1$. In practice,
$\sigma$ must be estimated. Since $T$ is usually very large in
typical applications, we shall for theoretical developments
treat $\sigma$ as known. Then, we can without loss of
generality set $\sigma= 1.$

For data involving $N$ sequences, to test the null hypothesis
$H_0$ that $\delta_i = 0$ for all $1 \leq i \leq N$ versus the
alternative $H_A$ that for some values of $\tau_1 < \tau_2$ at
least some $\delta_i$ are not zero, Zhang et al. \citep{Zhang2009b} proposed
a~direct generalization of (\ref{maxchisquare}):
%
\begin{equation}
\label{maxsumchi2} \max_{s < t} Z(s,t) \qquad \mbox{where }
Z(s,t)=\sum_{i = 1}^N U_i^2(s,t)
\end{equation}
and $U_i(s,t)$ is the sequence specific statistic
defined in (\ref{chgptstat}) for the $i$th sequence. Again, if
the variances are known, (\ref{maxsumchi2}) is the generalized
log likelihood ratio statistic for testing $H_0$ versus $H_A$.
For each fixed $s < t$, the null distribution of $Z(s,t)$ is
approximately $\chi^2$ with $N$ degrees of freedom. Even if
the
samples are related (say, replicates or members of the same family), this
relatedness only matters under the alternative
hypothesis that there is a~CNV. Thus, even for related
samples, as long as they are independent under the null hypothesis, the null
distribution of $Z(s,t)$ would be $\chi^2_N$. Large
values of $Z(s,t)$ are evidence against the null hypothesis. If
the null hypothesis is rejected, the maximum likelihood
estimate of the location of the variant interval is $(s^*,t^*)
= \argmax_{s,t} Z(s,t)$.

\subsection{Mixture model}
Whereas conducting a separate analysis for each individual
sequence requires that each sample show strong evidence for the
detection of a variant interval, the sum of $\chi^2$ statistic
goes to the other extreme of favoring situations where many samples have
relatively weak evidence. For cases where $N$ is moderately
large, say, in the 100s or even 1000s, it seems reasonable to
consider intermediate statistics that require each sample
to show moderate evidence before they are allowed to
make a~substantial contribution to the overall statistic.

Consider again the problem as originally formulated, where
$\mathcal{J}$ denotes the set of samples containing the same
variant interval, and let $Q_i(s,t)$ denote the indicator that
$i \in\mathcal{J}$ and that the aligned change-points are $s,t$.
If $Q_i(s,t)$ were observed, the generalized log-likelihood
ratio statistic, maximizing over the individual jumpsizes
$\{\delta_i\dvtx i=1,\dots,N\}$, would be
%
\begin{eqnarray}  \label{mixlikelihood}
&&\max_{s,t} \sum_{i=1}^N
\log\bigl[\{1-Q_i(s,t)\} + Q_i(s,t) e^{U^2_i(s,t)/2}\bigr] \nonumber\\[-8pt]\\[-8pt]
&&\qquad = \max_{s,t}
\sum_{i=1}^N Q_i(s,t) U^2_i(s,t)/2.\nonumber
\end{eqnarray}
Since $Q_i(s,t)$ is not observed, we have considered two
surrogate statistics. If we assume that $p_0 \in[0,1)$ is a
prior probability that $Q_i(s,t) = 1$, we could consider the
left-hand side of (\ref{mixlikelihood}) with $p_0$ substituted
for $Q_i (s,t)$, that is,
%
\begin{equation} \label{mixlr}
\max_{s,t} \sum_{i=1}^N
\log\bigl[1-p_0 + p_0 e^{U^2_i(s,t)/2}\bigr].
\end{equation}
This is the mixture likelihood ratio statistic. We could
also consider the posterior distribution of $Q_i(s,t)$, given
the data, which depends on the unknown parameters of the
problem. But if we maximize with respect to
these unknown parameters, we get
%
\begin{equation} \label{weighted} \max_{s,t} \sum_{i=1}^N
w_{p_0}[U^2_i(s,t)] U_i^2(s,t)/2,
\end{equation}
where
\begin{equation}\label{weightfun} w_{p_0}(x) = \exp(x/2)/\{r_{p_0} +
\exp(x/2)\},
\end{equation}
and $r_{p_0} = (1-p_0)/p_0$ denotes the prior odds against the
indicated hypothesis. We call this the weighted sum of chi-squares
statistic.

%
\begin{figure}

\includegraphics{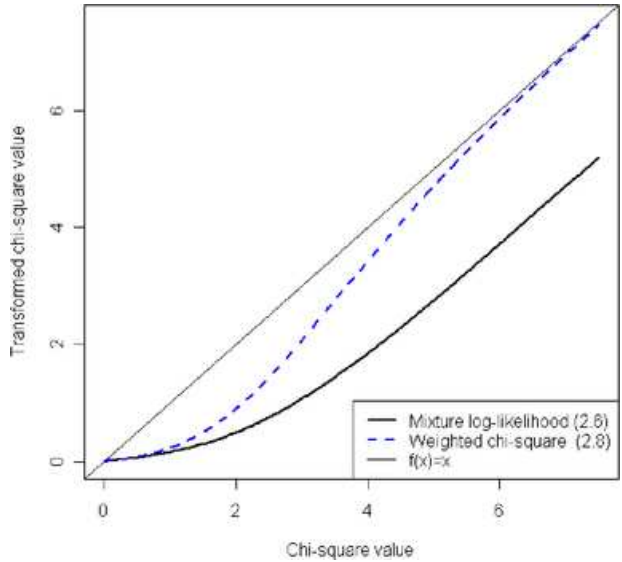}

\caption{Illustration of the transformations $U^2 \rightarrow\log(1-p_0+p_0
e^{U^2/2})$ (solid line) and $U^2 \rightarrow w_{p_0}(U^2)U^2/2$
(dashed line), with $p_0=0.1$.}\label{fig:2}
\end{figure}

Both the mixture likelihood ratio statistic and the weighted sum of the
chi-squares statistic are of the form of a maximum over $s<t$ of random
fields of the form $\sum_{i = 1}^N g[U_i(s,t)]$, where $g$ is a
suitable function. In Section~\ref{sec:tailapprox} we give an
approximation for the false positive rate of such a maximum for general
smooth functions $g$. The statistics we consider are all two-sided, and
can be considered to be transformations of the $\chi^2$ statistic
$U_i^2(s,t)$. The transformation $U^2 \rightarrow\log[(1-p_0)+p_0
\exp(U^2/2)]$ for the mixture likelihood and $U^2 \rightarrow
w_{p_0}(U^2)U^2$ for the weighted $\chi^2$ both effectively
soft-threshold the $\chi^2$ statistics, decreasing small values toward
zero. Figure \ref{fig:2} shows these transformations compared to the
identity transformation for the sum of chi-squares statistic. The new
statistics depend on the choice of $p_0$, with small values of $p_0$
requiring a more substantial apparent signal from a given sequence of
observations before that sequence is allowed to make an important
contribution to the overall statistic. For $p_0 = 1$, both recover the
sum of the chi-squares statistic. See Figure \ref{fig:2}.

\begin{remarks*}
(i) As we shall see in the power
analyses of Section \ref{sec:power}, these statistics are
relatively robust with respect to the choice of $p_0$.
Consequently, we have not considered an adaptive or data driven
method for estimating~$p_0$. (ii) Our original preference was
for the weighted sum of the chi-squares statistic, since the
heuristic argument behind this statistic suggests that it will
adapt better to the data than the mixture likelihood ratio. Our
numerical experiments indicate, however, that the two
statistics behave similarly, with the mixture likelihood ratio
being more stable and often slightly more powerful. Hence, we
report numerical results only for the mixture likelihood ratio
statistic.
\end{remarks*}

\section{Approximations for the significance level}\label{sec:tailapprox}
For scan statistics of the form
described above, we now give an analytic approximation for
the significance level that accounts for the simultaneous
testing of multiple dependent hypotheses.
The approximation gives a fast and computationally
simple way of controlling the false positive rate.

As described in Section \ref{sec:model}, we assume that the
data is a matrix of independent, identically distributed random
variables $y_{i,t}$ with mean zero, variance one and
sufficiently small tails. Each row represents a process and
there are~$N$ such processes. Given a starting point $s$ and an
interval length $\tau$, let $J_s^\tau= \{t\dvtx s < t \leq s+
\tau\}$ be a window of integers. Over this window construct,
for each process, the sum $W_{i}(J_s^\tau) = \sum_{t \in
J_s^\tau} y_{i,t}$ and consider the standardized statistic
\[
Z_i(J_s^\tau) = \tau^{-1/2}W_i(J_s^\tau) ,
\]
which again has
mean zero and variance one. Let $g$ be a smooth, positive
(nonlinear) function and consider the statistic $G(J_s^\tau) =
\sum_{i=1}^N g [Z_i(J_s^\tau) ]$. For example, $g(x) =
\log[(1-p_0)+p_0\exp(x^2/2)]$ for the mixture likelihood ratio
statistic. We are interested in the approximation of
%
\begin{equation} \label{probab}
\Prob \Bigl(\max_{s \leq
T}\max_{T_{0} \leq\tau \leq T_1} G(J_s^\tau)\geq x \Bigr)
\end{equation}
for $N$, $T_0,$ $T_1$ and $x$ diverging to $+\infty$
at the same rate.

In applying the above formulation to (\ref{mixlr}) and
(\ref{weighted}), we have already assumed $T$ is so large that the
standard deviations can be
estimated without error. To simplify the derivation,
we also assume the baseline mean values can be estimated
without error. At least for normally distributed variables
and $T_1 \ll T$ (the case of
interest here), this assumption does not change the
final approximation, and the required changes
are straightforward otherwise.
Hence, $\hat{\mu}_i$ and
$\hat{\sigma}_i$ are treated below as known constants and
$Z_i(J_s^\tau)$ is equivalent to $U_i(s,s+\tau)$.
When dealing with smaller (but still large) samples,
variation in the estimates of baseline parameters
can be handled
by modifications of the
same method.

To state our approximation, which involves an exponential change of
measure, we define the log moment generating function
\[
\psi_\tau(\theta) = \log\Exp\exp\{\theta g(Z_\tau)\},
\]
where $Z_\tau$
is a convenient notation for a random variable having the distribution of
the $Z_i(J_s^\tau)$. Now choose $\theta_\tau$ to satisfy
$\dot{\psi}_\tau(\theta_\tau) = {x/N}$, and let
%
\begin{equation} \label{eq:mean-sum-smooth} \mu(\theta) = \frac
{1}{2} \theta^2 \int[\dot g(z)]^2e^{\theta g(z) - \psi_\tau(\theta
)}\varphi(z)\,dz ,
\end{equation}
where
$ \varphi$ is the standard Gaussian density.

Then, provided that $T$ is subexponential in $N$, the
probability in (\ref{probab}) is asymptotically equivalent to
%
\begin{eqnarray} \label{approx}
&&\sum_{\tau=T_{0}}^{T_1}(T-\tau) e^{-N \{\theta_\tau\dot\psi
_\tau(\theta_\tau) - \psi_\tau(\theta_\tau)\}}
\{2 \pi N \ddot\psi_\tau(\theta_\tau)\}^{-1/2}
\nonumber\\[-8pt]\\[-8pt]
&&\hphantom{\sum_{\tau=T_{0}}^{T_1}}{}\times \theta
_\tau^{-1}
\mu^2(\theta_\tau)
(N/\tau)^2
\nu^2 \bigl([2 \mu(\theta_\tau)(N/\tau)]^{1 /2} \bigr) , \nonumber
\end{eqnarray}
where to a very good approximation
\[
\nu(x) \approx
[(2/x)\{\Phi(x/2)-1/2\}]/\{(x/2)\Phi(x/2) + \varphi(x/2)\}
\]
[cf. Siegmund and Yakir (\citeyear{SiegmundYakir2007})].
For the case of central interest in this paper, the $y_{i,j}$ are
standard normal, so $\psi_\tau$ does not depend on $\tau.$
Hence, several factors in (\ref{approx}) can be moved in front of
the sum; and the sum of the remaining terms can be approximated by an
integral, to obtain
%
\begin{eqnarray} \label{approxsimp}
&& N^2e^{-N \{\theta\dot\psi(\theta) - \psi(\theta)\}}
\{2 \pi N \ddot\psi(\theta)\}^{-1/2} \nonumber\\[-8pt]\\[-8pt]
&&\qquad {}\times \theta
^{-1} \mu^2(\theta)
\int_{T_0/T}^{T_1/T} \nu^2 \bigl([2N\mu(\theta)/(T t)]^{1/2}
\bigr)(1-t)\,dt/t^2.\nonumber
\end{eqnarray}

\begin{remark*}
(i) For the sum of the chi-squares statistic, $g(x) = x^2$, and~(\ref{approxsimp}) is
essentially the same as the approximation in
Zhang et al. \citep{Zhang2009b} except that $N -1$ has been replaced
by $N$ in two
places. (ii) Although the derivation
of (\ref{approxsimp}) requires that
$T_0 \rightarrow\infty,$ by an auxiliary argument one can show
in the normal case that
the approximation remains valid for arbitrarily small $T_0$,
in particular, for $T_0 = 1.$
\end{remark*}

%
\begin{table}
\caption{Accuracy of approximate thresholds:
The statistic is the mixture chi-square with
parameters $N = 100,  T_0 = 1,  T_1 = 50,   T = 500$.
The number of repetitions of the Monte Carlo experiment is 1000.
Results in parentheses are thresholds in units of
standard deviations above the mean}\label{table1}
\begin{tabular*}{\tablewidth}{@{\extracolsep{\fill}}lcd{2.7}c@{}}
\hline
$\bolds{p_0}$ & \textbf{Significance level} & \multicolumn{1}{c}{\textbf{Th (approx.)}} & \textbf{Th (MC)} \\
\hline
0.03 & 0.10 & 16.2 & 15.3 \\
0.03 & 0.05 & 17.1\ (8.7) & 16.8 \\
0.03 & 0.01 & 19.1 & 19.2 \\
0.1 & 0.10 & 27.4 & 26.3 \\
0.1 & 0.05 & 28.5\ (6.64) & 28.6 \\
0.1 & 0.01 & 30.9 & 31.3 \\
1.0 & 0.10 & 84.1 & 83.9 \\
1.0 & 0.05 & 85.9\ (5.08) & 85.8\\
1.0 & 0.01 & 89.8 & 99.8 \\
\hline
\end{tabular*}
\end{table}

\subsection{Accuracy of the approximation in the normal case}\label{sec3.1}

In this section we report a Monte Carlo experiment to verify
the accuracy of the suggested approximations for normally
distributed data. In Table \ref{table1} we consider the mixture likelihood ratio
and give significance thresholds based on simulation and on the
approximation (\ref{approxsimp}). It seems difficult to develop
intuition about the magnitude of these thresholds, so in a few
cases we have also included in parentheses the thresholds
measured in units of standard deviations above the mean. However, it
does not seem substantially easier
to develop intuition in this scale. The
corresponding threshold for a single normally distributed
sequence would be 4.3, so we see that in this scale the tail
of the distribution gets heavier with decreasing $p_0$, as one
would expect. While the
results in Table \ref{table1} indicate that the approximation is quite
accurate, the parameters~$N$,~$T_1$ and $T$ are all relatively
small, since the simulations become very time consuming for larger
values. A second example is given in the \hyperref[sec:proofs]{Appendix}.

\section{Power comparisons} \label{sec:power}

For the statistic $\max_{s, \tau} G(J_s^\tau)$, when the variant
interval is $(\tau_1, \tau_1 + \tau_2]$, we consider as an
approximation to the power of the probability
\[
\Prob\{ G(J_{\tau_1}^{\tau_2}) > b \},
\]
where $b$ is the threshold computed to achieve a pre-chosen
significance level, say, 0.05. This probability is a lower bound on the
true power, which also involves the much smaller probability that
$G(J_s^\tau) < b$ for $s = \tau_1,   \tau= \tau_2$, but exceeds
$b$ for nearby $s,\tau$. This simple approximation can be evaluated
using a~normal approximation or a small and fast Monte Carlo experiment
involving only $\tau_2\times N$ observations.

We conducted a power analysis for detecting CNV using the
Affymetrix 6.0 microarray platform, which contains $\sim$1.8
million probes. We assumed that a separate scan is conducted
for each chromosome. The average number of probes per
chromosome is around 80,000, and, thus, as a rough approximation,
we set the total length of a scan to be $T=80\mbox{,}000$. We
restricted our attention to the detection of short CNV, and,
thus, we enforced a maximum window size of $T_0=1000$. We
considered the detection of single copy insertions and
deletions, and assumed the signal to noise ratios (SNR) are
between 1 and 3. These are comparable to the signal to noise
ratios of actual data sets. For example, for the Hapmap data
set obtained from Affymetrix, we computed the signal to noise
ratios of those CNV detected in Zhang et al. \citep{Zhang2010} that are
confirmed by fosmid sequencing data. We found that the signal
to noise ratios for one copy gains are around 1.5--3.5 and that
for one copy losses are around 2--3.5. These SNR are higher than
true signal to noise ratios, since only those regions with
stronger signals were detected. The false positive rate is
controlled at $0.05/23 = 0.0022$, which corrects for the
multiple testing across chromosomes by the Bonferroni inequality.

\begin{figure}

\includegraphics{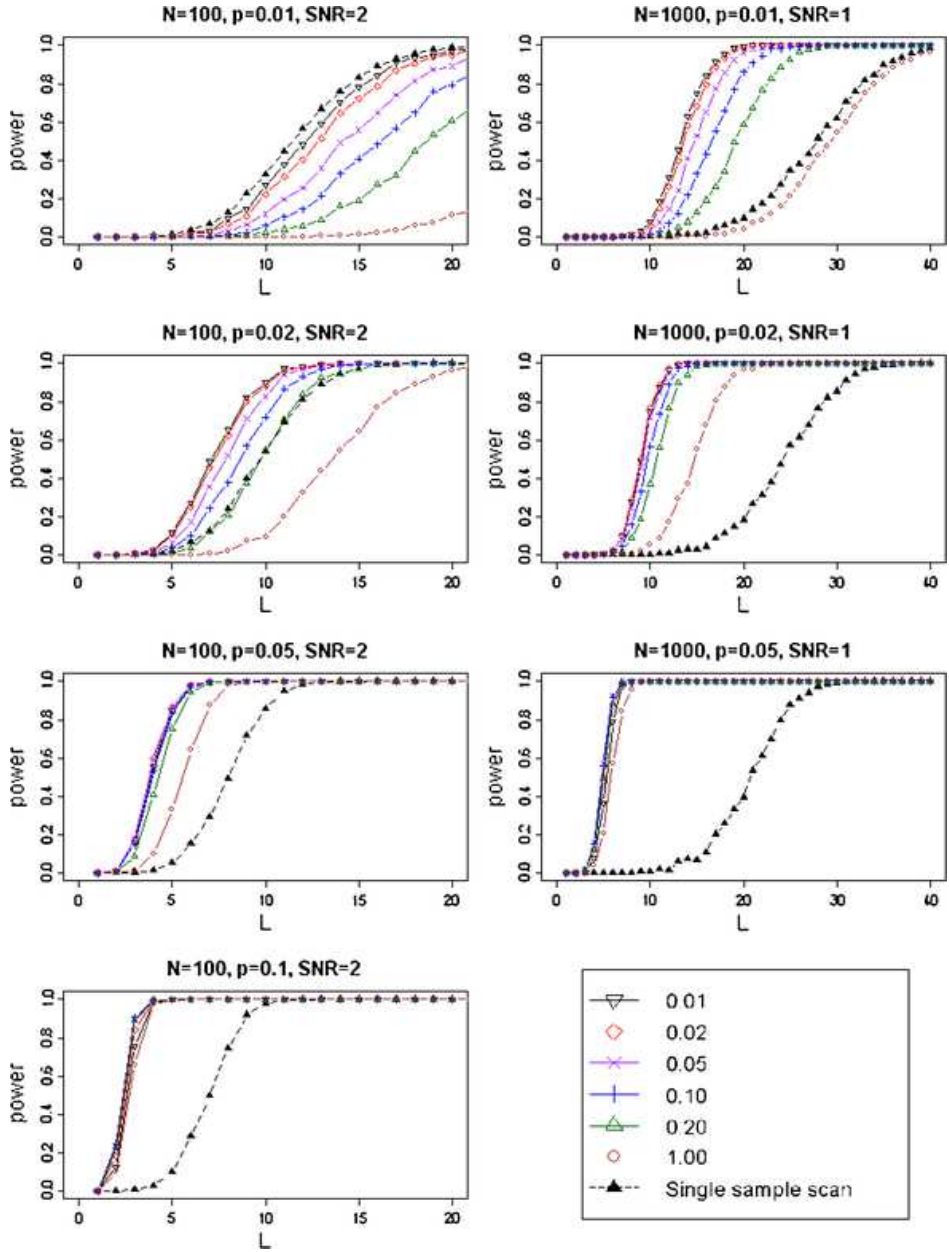}

\caption{Each plot shows the detection power versus length of
CNV for a given setting of sample size $N$, signal to noise
ratio (SNR) and fraction of carriers $p$. Going down each
column, $p$ increases while $N$ and SNR are fixed. $N=100$,
$\mathit{SNR}=2$ for the left column, and $N=1000$, $\mathit{SNR}=1$ for the
right column. The different curves represent the mixture scan
statistic (\protect\ref{mixlr}) for different values of $p_0$, with the
solid triangles representing the single sequence scan (see
legend at bottom right). \label{fig:power2}}
\end{figure}

Figure \ref{fig:power2} shows the power of detection of a CNV
of length $L$ that is present in a fraction
$p\in\{0.01,0.02,0.05,0.1\}$ of the cohort, using the scan
statistic~(\ref{mixlr}) with a range of values for $p_0$. The
size of the cohort $N$ is set to be 100 or 1000.
The signal to noise ratio is 2 in the left column, and 1 in the right
column. For each setting, Bonferroni corrected single-sample scans are
compared to multi-sample scans.

A few observations are worth noting from Figure
\ref{fig:power2}. First, when $N=100$ and $p=0.01$, that is,
when only one out of 100 samples carries a~change, a~single
sample scan has slightly greater power than a multi-sample scan using
a small value of $p_0$. In this case, using the sum of the chi-squares
statistic ($p_0=1$) can have very low power, which is expected. When the
signal is present in only one sample, pooling across samples should not
result in a~gain of power. When the true fraction $p$ is increased to
$0.02$, that is, only \emph{2 out of 100} samples carry a change, then a
multi-sample scan gives a~substantial boost in power for $p_0 \leq0.1$.
Furthermore, when the true fraction is $p=0.1$, a multi-sample
scan with any value of $p_0 \in(0.01,0.2)$ does better than a
single sample scan. These results also indicate that for $p$ not too small,
the results for different assumed values of $p_0$ are comparable.

%

Regarding the range of the horizontal axes in Figure \ref{fig:power2}, note
that for $N=100$ and signal to noise ratio of 2, the range of interval
lengths where we can expect a noticeable boost in power is typically less
than about 10--12. For
longer CNV, the power is already close to 1, so multi-sample scans do
not give added benefit. Note also that if the signal to noise ratio is
divided by $f$ and the length of the interval is multiplied by $f^{2}$,
the marginal power is unchanged. For example, if the signal to noise
ratio is
changed to 1, that is, $1/2$ as large, the noticeable boost in power occurs
for intervals up to four times as long, or about 40--50.

A surprising observation is that, for the range of
signal-to-noise ratios and interval lengths that seem relevant
to the current microarray platforms, scan statistics using a
small value of $p_0$ seem to be the winner under a wide range
of conditions. Even when the true fraction of carriers is a
moderate sized $p=0.1$, using $p_0=0.01$ gives almost the same
power as $p_0 = 0.1$ for most CNV
lengths. The benefit in using a large value of $p_0$ is more
noticeable when the signal to noise ratio is small while $N$ and
the percentage of carriers is
large, as expected.
(Results under a wider set of conditions are
available in supplementary materials.)

\section{Validation on a biological data set} \label{sec:data}

In Zhang et al. \citep{Zhang2009b} we illustrated our results on data
obtained with a set of 62 Illumina 550K Beadchips from
experiments performed on DNA samples extracted from
lymphoblastoid cell lines derived from healthy individuals.
These data were used recently as part of the quality assessment
panel at the Stanford Human Genome Center (i.e., they were
collected prior to studies of scientific interest to diagnose
possible problems in the experimental protocol). The 62 samples
are useful for method assessment because they represent 10 sets
of (child, parent, parent) trios and 16 technical replicates of
16 independent DNA samples. We withhold the relation between
samples during the detection process, so that the scanning
algorithm is blind to this information, and use it afterward
for validation. In Zhang et al. \citep{Zhang2009b} we used these data to
demonstrate the improvement of multi-sample scans based on the
sum of the chi-squares statistic over single sample scans. Here we
make a~similar comparison of the sum of the chi-squares statistic
with the mixture likelihood ratio statistic.

Data from most microarray based experiments exhibit various
artifacts, including strong local trends, first documented in
Olshen et al. \citep{Olshen2004} and studied in detail in Diskin et
al. \citep{Diskin2008}.
Diskin et al. \citep{Diskin2008} showed an association of these trends with
local GC content, and proposed a regression-based method that
reduced the magnitude of the local trends. Another problem
for microarray-based experiments is that the noise variance
varies significantly across probes, causing the bulk
distribution of the intensities for each sample to deviate from
normal. Such inhomogeneity of variances prompted
Purdom and Holmes \citep{PurdomHolmes} to use a Laplace distribution,
which can
be derived from a~mixture of normals with different variances,
to model gene expression data.

Before applying the cross sample scan, we preprocessed the
data so that the assumptions of independence and normality can
be valid. We adopted the following approach (let
$x=\{x_{it}\dvtx i=1,\dots,N; t=1,\dots,T\}$ be the raw data):
\begin{enumerate}
\item Each sample is standardized to its median, that is,
\[
x'_{it} = x_{it} -
\median(x_{it}\dvtx t=1,\dots,T).
\]
Let $x'$ be the matrix of $x'_{it}$ values obtained in this way.
\item Let $L$ be the rank-1 singular value
decomposition (SVD) of $x'$, and let
\[
x'' = x'-L.
\]
\item Standardize each SNP to have the same $84\%$ and
$16\%$ quantiles as the standard normal
distribution, that is,
\[
y_{it} = x_{it}''/d_t,
\]
where $d_t =(q_{t,84}-q_{t,16})/2$, where $q_{t,z}$
is the $z$th quantile of
$\{x_{it}''\dvtx i=1,\dots,N\}$.
\end{enumerate}
Empirically, we found that the rank-1 SVD of $x'$ in step 2
effectively captures experimental artifacts such as local
trends. This is because experimental artifacts can be viewed
as a low-rank perturbation of the data. For example,
Diskin et al. \citep{Diskin2008} showed that local trends can be
explained by
a linear model using one predominant factor, the local GC
content. In our data, we found that the rank-1 SVD can
eliminate local trends more completely than the genomic waves
software of Diskin et al., possibly because the local GC
content is not accurately computed or because local GC content
does not completely control for the artifacts. If the magnitude
of the signal (i.e., the CNV regions) is large compared to the
magnitude of artifacts, then parts of the signal would also be
captured by the SVD and dampened in step~2. However, in normal
DNA samples, the CNV regions are short and well separated.
Thus, compared to the sparse signal, artifacts overwhelmingly
contribute to the total data variation and almost completely
determine the rank-1 SVD. Finally, standardizing the quantiles
of each SNP in step 3 makes the assumption of normal errors
with homogeneous variance not too far from the truth.

%
\begin{figure}

\includegraphics{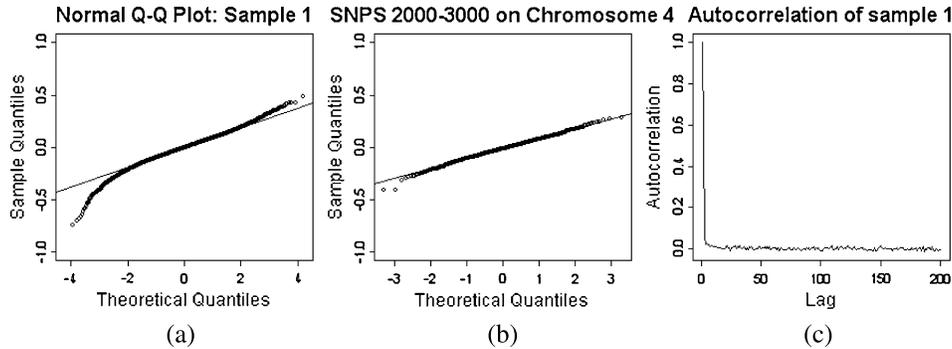}

\caption{Normal qq-plot [\textup{(a)} and \textup{(b)}] and autocorrelation plot \textup{(c)} for sample
1 of the Stanford Quality Control Panel data, after
preprocessing. The qq-plot in \textup{(a)} compares the distribution over all of
the SNPs (on all chromosomes)
for this sample against the standard normal distribution. The qq-plot
in \textup{(b)} zooms in on SNPs 2000--3000 on chromosome 4, which
does not contain any visually identifiable CNVs.} \label{fig:data2}
\end{figure}

Figure \ref{fig:data2} shows the normal qq-plot and the
autocorrelation plot for one of the 62 samples after this
normalization procedure. The qq-plot shows that the bulk of
the data now looks convincingly normal (the tails are heavier
than normal due to CNV regions), with the adherence to
normality more evident when we zoom in to a region that is
visually confirmed to contain no CNVs [Figure \ref{fig:data2}(b)].
From Figure~\ref{fig:data2}(c) we see that there is no detectable
autocorrelation in the normalized data.

To assess detection accuracy, we compare CNV identified for the
two technical replicates of the same individual, and also
compare those identified for the child with those identified
for the parents. We define ``inconsistency'' of detections of
CNV in individual samples as follows: (1) If a detected CNV in
one of the replicate pairs is not detected in the second sample
of the pair, the CNV is considered inconsistent. (2) If a
detected CNV in the child is not detected in at least one of
the parents, it is considered inconsistent. Detection accuracy
is thus assessed by plotting the number of
total versus inconsistent detections, and
different methods can be compared in such a~plot. See
Zhang et al. \citep{Zhang2009b} for a more complete discussion.

This method of accuracy assessment requires the
identification of the carriers of each CNV, and the method
of identification
affects the level of consistency. For
example, if all of the samples are classified as ``changed'' at
all CNV locations, then there would be many detections but no
inconsistencies.
In Zhang et al. \citep{Zhang2009b} we developed an empirically based
thresholding
method, which we use again here.

\begin{figure}

\includegraphics{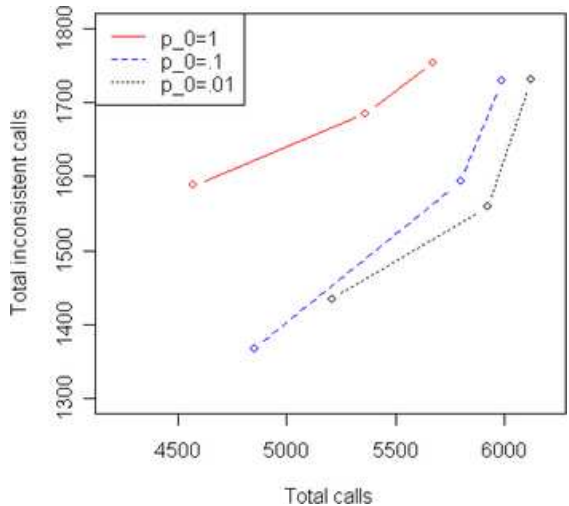}

\caption{Comparison of results on the Stanford Quality Control Panel
data using the mixture likelihood ratio statistic (\protect\ref{mixlr}). Each
curve is for a
different value of $p_0$. The points on the curve refer to different
absolute median thresholds $(0.2,0.3,0.4)$ for identifying carriers.}\label{fig:data}
\end{figure}

Figure \ref{fig:data} shows the results for different settings
of the parameters $p_0$ and the sample detection thresholds.
The horizontal axis is the number of total
detections and the vertical axis is the number of inconsistent
detections. Each line in the graph represents a different
setting for $p_0$, and dots on the line refer to performance at
varying values of a threshold parameter suggested in
Zhang et al. \citep{Zhang2009b}: $\delta_{\mathrm{MIN}}^{\mu}$, the absolute
difference in medians between the readings inside and outside
the interval for a sequence to be called a carrier of a CNV.
Within the range of $0.2$--$0.4$, as
$\delta_{\mathrm{MIN}}^{\mu}$ decreases, the size of the set declared to be
carriers, as well as the number of inconsistencies, increases.

%

Zhang et al. \citep{Zhang2009b} showed that using a multi-sample scanning
algorithm with the sum of the chi-squares statistic gives higher
consistency on these data than single sample analysis, and
Figure \ref{fig:data} shows that a mixture model, with small
values of $p_0$, gives an additional improvement. Using
$p_0=0.1$ performs noticeably better, and $p_0=0.01$ gives a
slight additional improvement.

Visual inspection of the data indicates that
most CNV regions are carried by fewer than 10 samples. Thus,
the fact that the mixture model with $p_0=0.01$ performs the
best is consistent with the power computations in Section~%
\ref{sec:power}. We also found that the detected CNV
region is often quite short. In many cases, consistent
calls contained fewer than 5 SNPs.

\section{Discussion}\label{sec6}

Although the scan statistic
relies on the unknown mixture fraction $p_0$, the power analyses show
that it is quite
insensitive to miss-specification of this parameter within
reasonable ranges.
Quite generally, the
power is sensitive to the value of $p_0$
only when $p$ or $p_0$ is very small. In practice, for $N = 1000$
it seems reasonable
to do a separate scan using a~few
different values of $p_0$, such as $p_0 \in\{0.001, 0.01, 0.1,
1\},$ and then apply a~Bonferroni correction.
One can also use a simple Monte Carlo approximation for the
marginal power as in
Section \ref{sec:power} to find a good range of $p_0$ to use
under various conditions.

From the power analysis in Section \ref{sec:power}, where we
assume the probe coverage and signal to noise ratios typical of
the Affymetrix 6.0 microarray platform (between two and three
standard deviations), we showed that the proposed method is
expected to boost power significantly for detection of short
CNV regions ($<$15 SNP coverage). When the signal to noise
ratio is weaker (around 1 standard deviation), we can expect an
improvement in power for CNV with less than 60 SNP coverage.
This is the range of CNV lengths where the current single
sample detection methods fail. In our experience such short
CNV are the most abundant in the genome, and would be the most
useful in a variety of studies. Many current genome-wide
studies simply ignore CNV with less than, say, 10 SNP coverage,
since they are not reliably detected with standard methods.
However, when we pool data across samples, the power increases
dramatically for the detection of such short CNV, even when
only a few samples within the cohort are carriers.

By assessing concordance across replicates and adherence to Mendelian
inheritance in parent--child trios, we showed in Section
\ref{sec:data} that the mixture likelihood ratio improves
the accuracy of CNV detection, especially when the variant is rare.
The accurate detection of rare variants makes these variants
available for genetic association studies and other studies
of population genetics.

The analytical approximation to the false positive rate given in
Section~\ref{sec:tailapprox} is accurate across all ranges of $N$ and
$p_0$ that we have tested. It allows instantaneous assessment of the
false positive rate for genome-wide scans, where Monte Carlo methods
are computationally infeasible. The theoretical framework for the
approximation is not limited to
Gaussian errors, and can be applied to other error models.

\begin{appendix}

\section{\texorpdfstring{Informal derivation of (\lowercase{\protect\ref{approx}})}{Appendix A: Informal derivation of (3.3)}}\label{sec:proofs}
The approximation (\ref{approx}) is obtained using a general method
for computing first passage probabilities first introduced in Yakir and
Pollack \citep{Yakir97} and further developed in Siegmund and Yakir
\citep{SiegmundYakir2000} and Siegmund, Yakir and Zhang \citep
{SiegmundYakirZhang}. The method relies on measure transformations that
shift the distribution for each sequence over the scan window. We use
the notation of Section \ref{sec:tailapprox}. We omit some of the technical details needed
to make the derivation rigorous. These details have been described and
proved in Siegmund, Yakir and Zhang \citep{SiegmundYakirZhang}.

Recall the definition $\psi_\tau(\theta)
= \log\Exp\exp\{\theta g(Z_\tau)\}$, where $Z_\tau$ is a
generic standardized sum over all observations within a given
window of size $\tau$ in one sample, and the parameter $\theta
= \theta_\tau$ is selected by solving the equation $ N \dot
\psi_\tau(\theta) = x$. Since $Z_{\tau}$ is a standardized sum
of $\tau$ independent random variables, $\psi_\tau$ converges
to a limit as $\tau\rightarrow\infty$, and $\theta_\tau$
converges to a limiting value. We denote this limiting value by
$\theta$. The transformed distribution for all sequences at
a fixed start position $s$ and window size $\tau$ is denoted by
$\Prob_s^\tau$ and is defined via
\[
{d}  \Prob_s^\tau= \exp[ \theta_{\tau} G(J_s^{\tau}) - N
\psi_\tau(\theta_\tau)] \,{d}  \Prob  .
\]
Let $\ell_N(J_s^{\tau}) = \log(d\Prob_s^\tau/d\Prob)$. Let $D =
\{(s,\tau)\dvtx 0<s<T, T_0\leq\tau\leq T_1\}$ be the set of all
possible windows in the scan. Let $A = \{\max_{(s,\tau) \in D}
G(J_s^\tau)\geq x\}$ be the event of interest. Then,
%
\begin{eqnarray}\label{proof1}
\Prob (A) &=&
\sum_{(s,\tau) \in D} \mathrm{E} \biggl[\exp[\ell_N(J_{s}^{\tau})]
\biggl(\sum_{(s',\tau')\in D} \exp[\ell_N(J_{s'}^{\tau'})]
\biggr)^{-1}; A \biggr]\nonumber\\
&=& \sum_{(s,\tau) \in D} \mathrm{E}_s^{\tau} \biggl[ \biggl(\sum_{(s',\tau
')\in D} \exp[\ell_N(J_{s'}^{\tau'})] \biggr)^{-1}; A
\biggr]\nonumber\\
&=&\sum_{(s,\tau) \in D} e^{\tell_N(J_s^{\tau})-\ell_N(J_s^{\tau
})}\nonumber\\[-8pt]\\[-8pt]
&&
{}  \times   \mathrm{E}_s^{\tau} \biggl[ \frac{\max_{u,v}
e^{\ell_N(J_u^v)-\ell_N(J_s^{\tau})}}{\sum_{u,v} e^{\ell
_N(J_u^v)-\ell_N(J_s^{\tau})}}e^{- \tell_N(J_s^\tau) - \log[\max
_{u,v}\ell_N(J_u^v)-\ell_N(J_s^{\tau})] }; A \biggr] \nonumber\\
&=& e^{-N \{\theta_\tau\dot\psi_\tau(\theta_\tau) - \psi_\tau
(\theta_\tau)\}}\nonumber\\
&&  {}\times\sum_{(s,\tau) \in D}
\mathrm{E}^{\tau}_s \biggl[\frac{ M_N(J_s^\tau)}{S_N(J_s^\tau)}\exp^{-\tell
_N(J_s^{\tau}) - \log M_N(J_s^{\tau})};A \biggr],\nonumber
\end{eqnarray}
where
\begin{eqnarray*}
\tell_N(J_s^\tau) &=& \sum_{i=1}^N \theta_\tau[g(Z_i(J_s^\tau)) -
\dot\psi_\tau(\theta_\tau)], \\
S_N(J_s^\tau) &=& \sum_{t,u} \exp \Biggl\{
\sum_{i=1}^N \theta_\tau[g(Z_i(J_t^u)) - g(Z_i(J_s^\tau))] \Biggr\},\\
M_N(J_s^\tau) &=& \max_{t, u} \exp \Biggl\{\sum_{i=1}^N \theta
_m[g(Z_i(J_t^u))
- g(Z_i(J_s^\tau))] \Biggr\}.
\end{eqnarray*}
Since $s$ and $\tau$ are fixed in much of what follows,
we sometimes suppress the dependence of the above
notation on $J_s^\tau$ and simply write $\tell_N, S_N, M_N$ for
$\tell_N(J_s^\tau), S_N(J_s^\tau),$ and $M_N(J_s^\tau)$,
respectively. As explained in Siegmund, Yakir and Zhang
(\citeyear{SiegmundYakirZhang}), under certain verifiable assumptions, a ``localization
lemma'' allows simplifying the quantities of the form
%
\begin{equation} \label{proof3}
\Exp_{s}^\tau [ (M_N / S_N )
e^{-\tell_N - \log M_N};
\tell_N + \log M_N \geq0  ]
\end{equation}
into much simpler expressions of the form
%
\begin{equation}\label{proof4}\sigma_{N,\tau}^{-1}(2\pi)^{-1/2}
\mathrm{E}[M/S],
\end{equation}
where $\sigma_{N,\tau}$ is the $\Prob_s^\tau$ standard deviation of
$\tell_N$ and $\mathrm{E}[M/S]$ is the limit of $\mathrm{E}[M_N/S_N]$ as
$N\rightarrow\infty$. This reduction relies on the fact that,
for large $N$ and~$T$, the ``local'' processes $M_N$ and $S_N$
are approximately independent of the ``global'' process
$\tell_N $. This allows the expectation in (\ref{proof3}) to
be decomposed into the expectation of $M_N/S_N$ times the
expectation involving $\tell_N+\log M_N$, treating $\log M_N$
essentially as a constant.

We next analyze each of the terms in (\ref{proof4}) separately.
First consider the processes $M_N$ and $S_N$. The difference
between standardized sums can be written in the form
\begin{eqnarray*}
Z_i(J_t^{u}) - Z_i(J_s^\tau) &=& Z_i(J_t^{u}) - u^{-1/2}W_i(J_s^\tau)
+ u^{-1/2}W_i(J_s^\tau) - Z_i(J_s^\tau)\\
&=& {u}^{-1/2} \bigl(W_i(J_t^u) - W_i(J_s^\tau) \bigr) + Z_i(J_s^\tau)
 [(\tau/u)^{1/2} - 1 ].
\end{eqnarray*}
By taking a Taylor expansion of order two and keeping only the
mean zero stochastic terms of order $O(N^{-1/2})$ and
deterministic terms of order $O(N^{-1})$, we obtain
\begin{eqnarray*}
g (Z_i(J_t^{u}) ) - g (Z_i(J_s^\tau) ) & \approx&
\frac{\dot g (Z_i(J_s^\tau) )}{u^{1/2}}
 \biggl(\sum_{j \in J_{t}^u\setminus J_s^{\tau}}     y_{i,j}
- \sum_{j \in J_{s}^\tau\setminus J_t^{u}}     y_{i,j} \biggr)\\
&&{} + Z_i(J_s^\tau)\dot g (Z_i(J_s^\tau) )
{\frac{\tau- u}{2u}}\\
&&{} + \frac{\ddot g (Z_i(J_s^\tau) )}{2u}
 \biggl(\sum_{j \in J_{t}^u\setminus J_s^{\tau}}     y_{i,j} ^2
+ \sum_{j \in J_{s}^\tau\setminus J_t^{u}}     y_{i,j}^2 \biggr) .
\end{eqnarray*}
It follows that
%
\begin{equation} \label{proof7}
\sum_{i=1}^{N} \theta_{\tau}  [g (Z_i(J_t^u) ) - g
(Z_i(J_s^\tau) )  ]
\approx\sum_{j \in J_{t}^u\setminus J_s^\tau}      \hat H_{j}^+
+ \sum_{j \in J_{s}^\tau\setminus J_t^{u}}      \hat H_j^-
\end{equation}
for
\begin{eqnarray*}
\hat H_{j}^+ &=& \frac{\theta_\tau N^{1/2}}{u^{1/2}}
 \Biggl(N^{-1/2}\sum_{i=1}^{N} \dot g (Z_i(J_s^\tau) )
y_{i,j} \Biggr)\\
&&{}+ \frac{\theta_\tau N }{2u}
 \Biggl({N}^{-1}\sum_{i=1}^{N} [ \ddot g (Z_i(J_s^\tau) )
y_{i,j} ^2 - Z_i(J_s^\tau)\dot g (Z_i(J_s^\tau) ) ]
\Biggr),\\
\hat H_{j}^- &=& \frac{-\theta_\tau N^{1/2}}{u^{1/2}}
 \Biggl(N^{-1/2}\sum_{i=1}^{N} \dot g (Z_i(J_s^\tau) )
y_{i,j} \Biggr)\\
&&{} + \frac{\theta_\tau N}{2u}
 \Biggl(N^{-1}\sum_{i=1}^{N}  [\ddot g (Z_i(J_s^\tau) )
y_{i,j} ^2 + Z_i(J_s^\tau)\dot g (Z_i(J_s^\tau) ) ]
\Biggr) .
\end{eqnarray*}
Observe that one may substitute $\tau$ for $u$ and $\theta=
\lim_{\tau\rightarrow\infty} \theta_\tau$ for $\theta_\tau$
in the definition of the increments and still maintain the
required level of accuracy.

Consider the random variable $\hat H_{j}^+$. Its first
component has mean zero under the distribution determined by
$\Prob_s^\tau$, since the random variables $y_{i,j}$, $1 \leq i
\leq N$, are not in the interval $J_s^\tau$. By the central
limit theorem, $\hat H_{j}^+$~converges to a normal random
variable with variance that is approximately equal to
\[
\Var_s^\tau  [\hat H_{j}^+ ] \approx\theta^2 \frac
{N}{\tau}
\Var_s^\tau  ( \dot g (Z_1(J_s^\tau) ) y_{1,j} )
\approx\theta^2 \frac{N}{\tau}
\Expec_\theta   [ \dot g(Z)^2  ] ,
\]
with the $\Prob_\theta$ distribution of the random variable $Z$
given by a density proportional to $\varphi(z) e^{\theta
g(z)}$, for $\theta$ the limit of $\theta_\tau$. The second
component converges by the law of large numbers to
\[
\Expec_s^\tau   [\hat H_{j}^+ ] \approx\frac{\theta
N}{2\tau}
\Expec_\theta  [ \ddot g(Z)
- \dot g(Z) Z ] = -\frac{1}{2}\theta^2\frac{ N}{\tau}
\Expec_\theta  [( \dot g(Z))^2 ] ,
\]
where the last equation follows from integration of the
identity
\[
\frac{d}{dz} \bigl[\varphi(z)\dot g(z) e^{\theta g(z)} \bigr] =
[-z \dot g(z) + \ddot g(z)
+ \theta(\dot g(z))^2  ] \varphi(z)e^{\theta g(z)}\,dz .
\]

Regarding the random variable $\hat H_{j}^-$, note that due to
the sufficiency of the statistic $Z_i(J_s^\tau)$ and the
exchangeability of the observations that form it under the null
distribution, we get that the conditional expectation of~%
$y_{i,j}$, given $Z_i(J_s^\tau)$, equals
$Z_i(J_s^\tau)/\sqrt{\tau}$. Straightforward computations, that
essentially repeat those carried out for $\hat H_j^+$, show
that
\[
\Expec_s^\tau   [\hat H_{j}^- ] \approx-\frac{1}{2}\theta^2
\frac{N}{\tau}
\Expec_\theta  [( \dot g(Z))^2 ] .
\]
For the variance of this term, since one can ignore $o(1)$ quantities,
we should approximate the expectation
\[
\Expec_s^\tau   \{ [\dot g (Z_1(J_s^\tau) ) ]^2
y_{1,j}^2 \} .
\]
But, if we denote by $\tilde Z= Z_1(J_s^\tau\setminus\{j\})$
the standardized sum of all the observations in the first row
excluding $y_{1,j}$, and denote by $\tilde\Expec_s^\tau$ the
expectation with respect to the measure where $g(\tilde Z)$ is
used for the exponential change of measure, we get a negligible difference
between the original expectation and
\begin{eqnarray*}
\tilde\Expec_s^\tau   \{ [\dot g (\tilde Z ) ]^2
y_{1,j}^2 \}
&=& \tilde\Expec_s^\tau   \{ [\dot g (\tilde Z )
]^2  \}\\
&\approx&\Expec_\theta  [( \dot g(Z))^2 ] .
\end{eqnarray*}
The difference is negligible due to the facts that the function
$h(z,\theta) = [\dot g(z)] e^{\theta g(z)}$
is continuous with respect to both $z$ and $\theta$ and
that $\psi_\tau$ converges,
as $\tau\rightarrow\infty$ to a continuous limit.
The conclusion is that both types of increments converge to the
same limiting normal distribution,
with a mean value equal to minus one half the variance.

One may use the same technique in order to show that the
covariance between any two increments is of the order of
$O(1/N)$.

The process $\tell_N$ has mean 0 and variance
%
\begin{eqnarray}  \label{proof6}
\sigma^2_{N,\tau} &=&\Var_s^\tau(\tell_N)\nonumber \\
& =& N \theta_\tau^2 \ddot
\psi_\tau(\theta_\tau) \\
&=& N
\theta_\tau^2
\Var_s^\tau(g(Z_1(J_s^\tau))\nonumber
\end{eqnarray}
and its covariance with an increment of the local process is of order
$N^{-1/2}$, so asymptotically the two are independent.

It follows from these calculations that the two local processes
in (\ref{proof7}) which arise from perturbations at the
endpoints of the interval $(s, s+\tau]$ are asymptotically
independent two-sided random walks. The increments are
independent, identically distributed normal random variables.
Moreover, integrating by parts the analytic expression for
$\mathrm{E}_s^\tau[\ddot g( Z_{i,s})]$, one sees that the absolute
value of the mean of the local process equals half the
variance. The random variables $M_N$ and $S_N$ are
respectively the maximum and sum of these local processes.
Consequently, following Siegmund and Yakir \citep{SiegmundYakir2000},
we get that
%
\begin{equation}\label{proof5}\Expec[\mathcal{M}/\mathcal{S}] =
\bigl[(N/\tau) \mu(\theta)
\nu \bigl([2(N/\tau) \mu(\theta)]^{1 /2} \bigr)\bigr]^2,
\end{equation}
where
\begin{eqnarray*}
\mu(\theta) &=& \frac{\theta^2}{2}
\Expec_\theta  [ \{\dot g(Z)\}^2 ] \\
&=& \frac{\theta^2}{2}
\int[\dot g(z)]^2e^{\theta g(z) - \psi(\theta)}\varphi(z)\,dz .
\end{eqnarray*}
Combining (\ref{proof5}) with (\ref{proof6}) in (\ref{proof4}),
and then substituting the result for the expectations in
(\ref{proof1}) yields (\ref{approx}).

\section{\texorpdfstring{Another numerical example}{Appendix B: Another numerical example}}

The numerical example discussed in Section \ref{sec3.1} was limited to relatively
small values of $T$ by the extremely time consuming nature of the
simulations. Here we give a somewhat different example where it is
computationally feasible to consider larger $T$, since the scan
statistic involves only a one-dimensional maximization.

The statistic is
%
\begin{equation} \label{link}
\max_{0 < j \Delta< \ell} \sum_{i = 1}^N \log\bigl[1-p_0 + p_0
\exp\bigl(U^2_{i}( j \Delta)/2\bigr)\bigr],
\end{equation}
where the processes $U_i(t)$ are independent stationary
Ornstein--Uhlenbeck proc-esses with covariance function
$\Cov[U_i(s), U_i(t)] = \exp(-\beta|t-s|).$ This statistic
would be reasonable as an approximation in a linkage study of
the expression levels of $N$ genes, regarded as quantitative
traits (eQTL), when one is particularly interested in ``master
regulators,'' that is, genomic regions that control the expression
levels of a collection of genes. See
Siegmund and Yakir \citep{SiegmundYakir2007} for a general discussion
of linkage analysis and Morley et al. \citep{Morley2004}, G\"oring et
al. \citep{Goring2007} and Shi, Siegmund and Levinson \citep{Shi2007}
for recent studies of linkage for eQTL and discussions of the existence
of master regulators. In this case $\ell$ is the length of the genome
in centimorgans (taken here to be 1600, the approximate length of a
mouse genome),
$\Delta$ is the (average) genetic distance between markers, and $\beta
= 0.02$ for a backcross or for the statistic associated
with the additive effect of an intercross. Table~\ref{table2}
gives numerical results for an approximation to the
tail probability of (\ref{link}), which was suggested by
Siegmund, Yakir and Zhang (\citeyear{SiegmundYakirZhang}) and is analogous to
(\ref{approx}), but is much simpler to derive. This approximation
is also quite accurate.
%
\begin{table}
\tablewidth=270pt
\caption{Accuracy of approximate thresholds:
The statistic is the mixture likelihood ratio for linkage, with
parameters $N = 1000,  \ell= 1600 ,   \Delta= 1,   \beta= 0.02$.
The number of repetitions of the Monte Carlo experiment is 1000}\label{table2}
\begin{tabular*}{\tablewidth}{@{\extracolsep{\fill}}lccc@{}}
\hline
$\bolds{p_0}$ & \textbf{Significance level} & \textbf{Th (approx.)} & \textbf{Th (MC)} \\
\hline
0.02 & 0.10 & 47.0 & 47.5 \\
0.02 & 0.05 & 48.5 & 48.9 \\
0.02 & 0.01 & 51.3 & 51.8 \\
0.01 & 0.10 & 30.1 & 29.2 \\
0.01 & 0.05 & 31.3 & 31.5 \\
0.01 & 0.01 & 33.6 & 33.8 \\
\hline
\end{tabular*}
\end{table}

\end{appendix}

\printaddresses

\end{document}